\documentclass[conference]{IEEEtran}
\IEEEoverridecommandlockouts
\usepackage{cite}
\usepackage{amsmath,amssymb,amsfonts}
\usepackage{algorithmic}
\usepackage{graphicx}
\usepackage{textcomp}
\usepackage[dvipsnames]{xcolor}
\usepackage{multirow}
\usepackage{makecell}
\usepackage{siunitx}  % https://tex.stackexchange.com/questions/2746/
\usepackage{caption}
\usepackage{subcaption}
\usepackage[hidelinks]{hyperref}
\usepackage{cleveref}
\usepackage[indentafter]{titlesec}
\usepackage{setspace}

\graphicspath{{figures/}}

\def\BibTeX{{\rm B\kern-.05em{\sc i\kern-.025em b}\kern-.08em
    T\kern-.1667em\lower.7ex\hbox{E}\kern-.125emX}}

\def\baselinestretch{0.98}

\makeatletter

\titlespacing*{\section}{0pt}{0.8\baselineskip minus 0.5\baselineskip}{0.1\baselineskip}
\titlespacing*{\subsection}{0pt}{0.8\baselineskip minus 0.5\baselineskip}{0.1\baselineskip}
\titleformat{\subsubsection}[runin]{\itshape}{\arabic{subsubsection})}{0.5em}{}[:]
\titlespacing*{\subsubsection}{\parindent}{0.1\baselineskip}{0.5em}

\g@addto@macro\normalsize{%
  \setlength\abovedisplayskip{3pt plus 2pt minus 1pt}
  \setlength\belowdisplayskip{3pt plus 2pt minus 1pt}
  \setlength\abovedisplayshortskip{2pt plus 2pt minus 1pt}
  \setlength\belowdisplayshortskip{2pt plus 2pt minus 1pt}
}

\makeatother

\begin{document}

\title{The Conformer Encoder May Reverse \\ the Time Dimension}

% authors: robin, albert, mohammad, ralf, ney

\author{
\IEEEauthorblockN{
	Robin Schmitt, 
	Albert Zeyer, 
	Mohammad Zeineldeen, 
	Ralf Schlüter and 
	Hermann Ney
	}
\IEEEauthorblockA{Human Language Technology and Pattern Recognition, Computer Science Department,\\RWTH Aachen University, Aachen, Germany}
\IEEEauthorblockA{AppTek GmbH, Aachen, Germany}
\IEEEauthorblockA{Email: robin.schmitt1@rwth-aachen.de, \{zeyer,zeineldeen,schlueter,ney\}@cs.rwth-aachen.de}
}

\maketitle

% don't put the figure to the left column of the first page
% https://tex.stackexchange.com/questions/63131/start-placing-figures-on-right-hand-side-column-of-first-page
\global\csname @topnum\endcsname 0
\global\csname @botnum\endcsname 0

% approximately 100 to 150 words
\begin{abstract}

We sometimes observe monotonically \emph{decreasing} cross-attention weights
in our Conformer-based global attention-based encoder-decoder (AED) models,
% addition from rebuttal
%\textcolor{Green}{
negatively affecting performance compared to monotonically 
increasing attention weights.
%}
Further investigation shows that the Conformer encoder reverses the sequence in the time dimension.
We analyze the initial behavior of the decoder cross-attention mechanism 
and find that it encourages the Conformer encoder self-attention
to build a connection between the initial frames
and all other informative frames.
Furthermore, we show that, at some point in training,
the self-attention module of the Conformer
starts dominating the output over the preceding feed-forward module,
which then only allows the reversed information to pass through.
We propose methods and ideas of how this flipping can be avoided and 
investigate a novel method to obtain label-frame-position alignments
by using the gradients of the label log probabilities w.r.t.~the encoder input frames.

\end{abstract}


\section{Introduction \& related work}

% \begin{figure}[b]
% 	\centering
% 	\includegraphics[width = 0.9\columnwidth]{att-weights/multi-gpu/conformer-rand-4321_ep100/plot.train-other-960_1578-6379-0013_1578-6379-0013.pdf}
% 	\caption{Cross-attention weights of model with reversed encoder.}
% 	\label{fig:flipped-att-weights}
% \end{figure}

% \begin{figure}
% 	\centering
% 	\includegraphics[width = 0.9\columnwidth]{att-weights/multi-gpu/conformer-w-ctc_ep350/plot.train-other-960_1578-6379-0013_1578-6379-0013.pdf}
% 	\caption{Cross-attention weights of model with normal encoder.}
% 	\label{fig:non-flipped-att-weights}
% \end{figure}

{
\setlength{\belowcaptionskip}{-1pt}
\begin{figure}
	\centering
	\includegraphics[width = 1.\columnwidth]{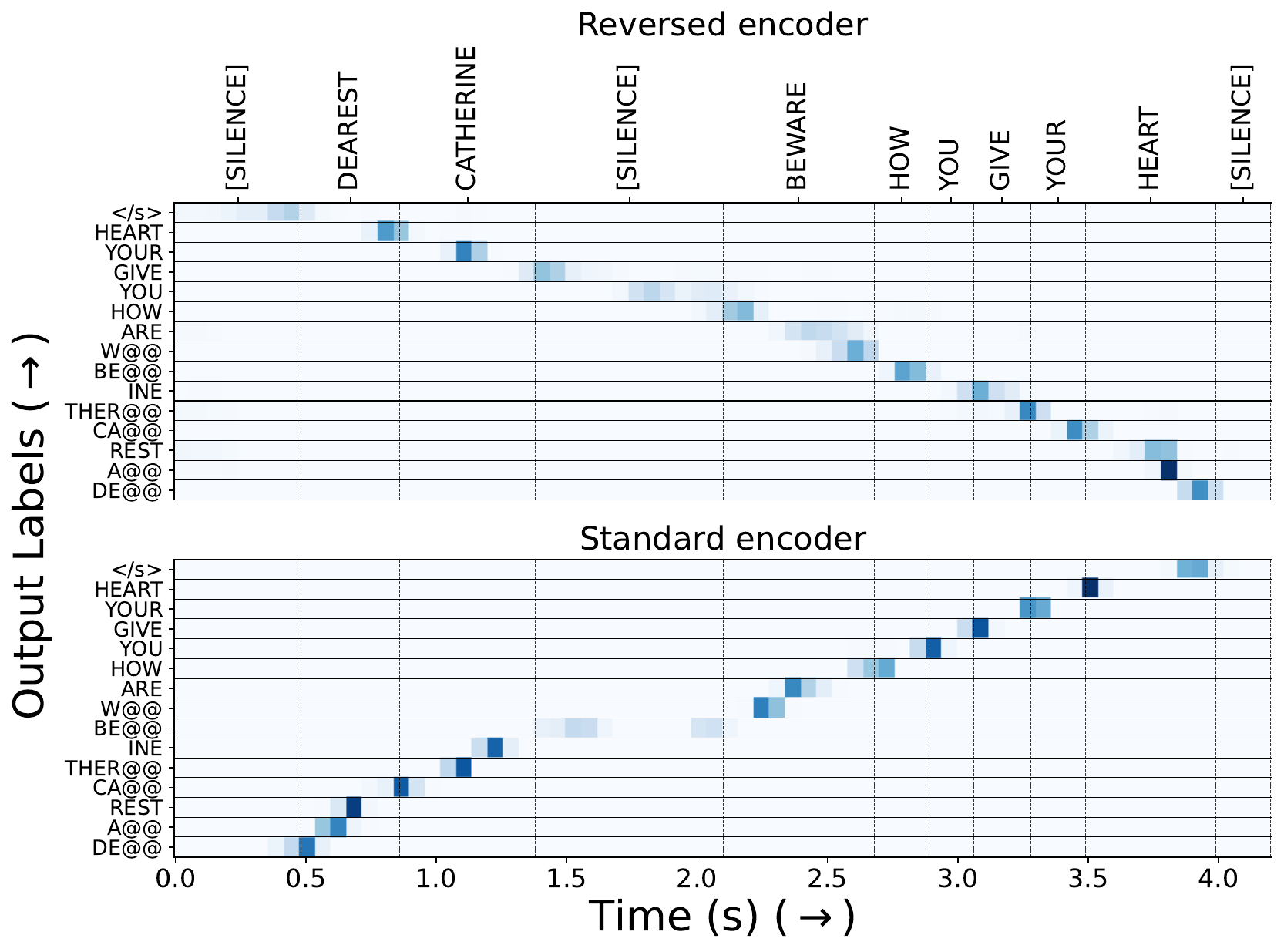}
	\caption{Cross-attention weights of a model with reversed encoder vs. standard encoder.}
	\label{fig:non-flipped-vs-flipped-att-weights}
\end{figure}
}

The Conformer encoder \cite{gulati2020conformer}
is widely used in automatic speech recognition (ASR) systems.
It combines self-attention and convolutional layers
to model both long-range and local dependencies in the input sequence
and outperforms \cite{gulati2020conformer,guo2020espnet-conformer}
the Transformer encoder \cite{vaswani2017transformer}.
In the attention-based encoder-decoder (AED) framework, the decoder
autoregressively produces an output sequence while attending to the whole
encoder output at each step
\cite{chorowski2015asr-aed,chan2016las}.

We observe that the Conformer encoder of an AED model
sometimes \emph{reverses the time dimension} of the input sequence,
as the cross-attention weights show (\Cref{fig:non-flipped-vs-flipped-att-weights}).
This work investigates why this phenomenon occurs
and how it can be avoided.

Furthermore, we propose a novel way to obtain label-frame-position alignments
(for each output label, the start/end time frames in the input)
by leveraging the gradients of the label log probabilities w.r.t.~the encoder input frames.
We use methods related to saliency maps and
other gradient-based attribution methods such as the ones studied in \cite{ancona2017gradient-based-attribution,prasad2020accents-gradients}.
Moreover, \cite{SerranoS19,Hechtlinger16} shows how gradients could be used to better interpret the behavior of a model.

% those:
% https://proceedings.neurips.cc/paper_files/paper/2021/hash/ba3c736667394d5082f86f28aef38107-Abstract.html Understanding How Encoder-Decoder Architectures Attend (edited) 
% https://arxiv.org/abs/1906.03731 Is Attention Interpretable? (edited) 
% https://arxiv.org/abs/1611.07634 Interpretation of Prediction Models Using the Input Gradient (edited) 
% https://arxiv.org/abs/1902.10186 Attention is not Explanation
% https://arxiv.org/abs/1703.03717 Right for the Right Reasons: Training Differentiable Models by Constraining their Explanations
% https://arxiv.org/abs/1711.06104 Towards better understanding of gradient-based attribution methods for Deep Neural Networks
% https://arxiv.org/abs/2202.00673 Visualizing Automatic Speech Recognition -- Means for a Better Understanding?

\section{AED Model}

Our baseline is the standard AED model adapted for ASR \cite{chorowski2015asr-aed,chan2016las}.
The encoder consists of a convolutional frontend,
which downsamples the sequence $x_1^{T'}$ of 10ms frames
into a sequence ${h_0}_1^T$ of 60ms frames,
followed by a stack of $N=12$ Conformer blocks,
which further process the downsampled sequence
resulting in the encoder output $h_1^T$:
\begin{align*}
	{h_0}_1^T &= \operatorname{ConvFrontend}(x_1^{T'}), \\
	{h_i}_1^T &= \operatorname{ConformerBlock}_i ({h_{i-1}}_1^T), \quad h = h_N, i=1,\dots,N .
\end{align*}
The probability of the output label sequence $a_1^S$ given the encoder output $h_1^T$ is defined as
\begin{equation*}
	p(a_1^S \mid h_1^T) = \prod_{s=1}^S p(a_s \mid a_1^{s-1}, h_1^T).
\end{equation*}
\newcommand{\eos}{\ensuremath{\texttt{EOS}}}
At each step $s$, the decoder autoregressively predicts the next label $a_s$ by attending over the whole encoder output $h_1^T$.
We define $a_S=\eos{}$ as the end-of-sequence token,
which implicitly models the probability of the sequence length.
We use an LSTM \cite{hochreiter1997lstm} decoder with single-headed MLP cross-attention \cite{LuongPM15}
following our earlier setup \cite{zeyer2018interspeech}.
We use BPE subword label units \cite{Sennrich:bpe} (1k and 10k vocab.~sizes).

We train our model using the standard label-wise cross-entropy (CE) criterion
$L = -\log p(\overline{a}_1^S \mid h_1^T)$
using the target transcriptions $\overline{a}_1^S$.
In the experiments, where we observe the flipping of cross-attention weights,
there is no further loss,
but we also tested to add CTC \cite{graves2006ctc} as an auxiliary loss \cite{hori2017attctc}.

We use label-synchronous beam search for recognition.

\section{Experimental Setup}

Our experimental setup follows the global AED baseline from \cite{zeineldeen2024:ChunkedAED}.
We perform experiments on the LibriSpeech 960h \cite{panayotov2015librispeech} corpus
using the RETURNN framework \cite{zeyer2018returnn}
based on PyTorch \cite{paszke2019pytorch}.
Our pipeline is managed by Sisyphus \cite{PeterBN18}.
All models are trained for 100 epochs with the AdamW optimizer \cite{loshchilov2017adamw},
using on-the-fly speed perturbation and SpecAugment \cite{park2019specaugment}.
As hardware, we either use a single Nvidia A10 GPU or 4x Nvidia 1080 GTX GPUs in parallel
with parameter syncing every 100 batches
\cite{McDonald-2010-DistributedTrainingStrategies,Zhang-2014-ImprovingDeepNeural}.
All the code to reproduce our experiments is published%
\footnote{\scriptsize\url{https://github.com/rwth-i6/returnn-experiments/tree/master/2024-flipped-conformer}}.

\begin{table}
	\centering
	\caption{Comparing the baseline in this paper -- Conformer AED with BPE 1k without CTC aux.~loss
	-- vs related models.
	Results on LibriSpeech, without external language model.}
	\label{tab:baseline-wers}
	\setlength{\tabcolsep}{3pt}
	\footnotesize
    \begin{tabular}{|l|c|c|c|c|cccc|}
        \hline
        \multirow{3}{*}{AED Model}              & \multirow{3}{*}{\begin{tabular}[c]{@{}c@{}}Label \\ Units\end{tabular}} & \multirow{3}{*}{\begin{tabular}[c]{@{}c@{}}CTC\\ aux.\\ loss\end{tabular}} & \multirow{3}{*}{\begin{tabular}[c]{@{}c@{}}Dis.\\ self-att.\\ 1st ep.\end{tabular}} & \multirow{3}{*}{\makecell{Flip. \\ enc.}} & \multicolumn{4}{c|}{WER [\%]}                                                                \\ \cline{6-9} 
                                                &                                                                         &                                                                            &                                                                                          &                          & \multicolumn{2}{c|}{dev}                                & \multicolumn{2}{c|}{test}          \\ \cline{6-9} 
                                                &                                                                         &                                                                            &                                                                                          &                          & \multicolumn{1}{c|}{clean} & \multicolumn{1}{c|}{other} & \multicolumn{1}{c|}{clean} & other \\ \hline
        \makecell{E-Branchf.-\\Trafo \cite{KimWPPSHW22}} & BPE 5k                                                                  & Yes                                                                        & No                                                                                       & No                       & \multicolumn{1}{c|}{2.0}   & \multicolumn{1}{c|}{4.6}   & \multicolumn{1}{c|}{2.1}   & 4.6   \\ \hline \hline
        \multirow{7}{*}{\makecell{Conformer-\\LSTM}}         & \multirow{5}{*}{BPE1k}                                                  & \multirow{2}{*}{No}                                                        & \multirow{3}{*}{No}                                                                      & Yes                      & \multicolumn{1}{c|}{2.8}   & \multicolumn{1}{c|}{6.8}   & \multicolumn{1}{c|}{3.0}   & 6.6   \\ \cline{5-9} 
                                                &                                                                         &                                                                            &                                                                                          & \multirow{6}{*}{No}      & \multicolumn{1}{c|}{2.3}   & \multicolumn{1}{c|}{5.8}   & \multicolumn{1}{c|}{2.5}   & 5.8   \\ \cline{3-3} \cline{6-9} 
                                                &                                                                         & Yes                                                                        &                                                                                          &                          & \multicolumn{1}{c|}{2.2}   & \multicolumn{1}{c|}{5.6}   & \multicolumn{1}{c|}{2.5}   & 5.6   \\ \cline{3-4} \cline{6-9} 
                                                &                                                                         & No                                                                         & \multirow{2}{*}{Yes}                                                                     &                          & \multicolumn{1}{c|}{2.2}   & \multicolumn{1}{c|}{5.4}   & \multicolumn{1}{c|}{2.5}   & 5.7   \\ \cline{3-3} \cline{6-9} 
                                                &                                                                         & Yes                                                                        &                                                                                          &                          & \multicolumn{1}{c|}{2.3}   & \multicolumn{1}{c|}{5.7}   & \multicolumn{1}{c|}{2.5}   & 5.5   \\ \cline{2-4} \cline{6-9} 
                                                & \multirow{2}{*}{BPE10k}                                                 & No                                                                         & \multirow{2}{*}{No}                                                                      &                          & \multicolumn{1}{c|}{2.6}   & \multicolumn{1}{c|}{6.1}   & \multicolumn{1}{c|}{2.8}   & 6.0   \\ \cline{3-3} \cline{6-9} 
                                                &                                                                         & Yes                                                                        &                                                                                          &                          & \multicolumn{1}{c|}{2.3}   & \multicolumn{1}{c|}{5.6}   & \multicolumn{1}{c|}{2.5}   & 5.7   \\ \hline
        \end{tabular}
\end{table}

\Cref{tab:baseline-wers} compares the word error rate (WERs) of our baseline model
to related models.
BPE 1k without CTC aux.~loss is the base configuration of all experiments.
% addition from rebuttal
%\textcolor{Green}{
Flipping has a
negative effect on the WER while our measures against flipping have a positive 
effect. Notably, our flipped models show significant variation in WER
(6.8-21.6\% on dev-other) when using different random seeds, with only the 
best result included in the table.
%}

\section{Analysis}

% just write down some thoughts for now to have some initial structure

\subsection{Initial Development of Cross-Attention Weights}
\label{sec:initial-dev-cross-att}

% \begin{figure}
% 	\centering
% 	\begin{subfigure}[b]{0.5\textwidth}
% 	   \includegraphics[width=1\linewidth]{att-weights/development/conformer-rand-43321/ep10_train-other-960_1578-6379-0013_1578-6379-0013.pdf}
% 	%    \caption{}
% 	   \label{fig:Ng1} 
% 	\end{subfigure}

% 	\vspace{-.4cm}

% 	\begin{subfigure}[b]{0.5\textwidth}
% 		\includegraphics[width=1\linewidth]{att-weights/development/conformer-rand-43321/ep40_train-other-960_1578-6379-0013_1578-6379-0013.pdf}
% 	 %    \caption{}
% 		\label{fig:Ng2}
% 	 \end{subfigure}

% 	 \vspace{-.4cm}

% 	 \begin{subfigure}[b]{0.5\textwidth}
% 		\includegraphics[width=1\linewidth]{att-weights/development/conformer-rand-43321/ep50_train-other-960_1578-6379-0013_1578-6379-0013.pdf}
% 	 %    \caption{}
% 		\label{fig:Ng2}
% 	 \end{subfigure}

% 	 \vspace{-.4cm}

% 	 \begin{subfigure}[b]{0.5\textwidth}
% 		\includegraphics[width=1\linewidth]{att-weights/development/conformer-rand-43321/ep60_train-other-960_1578-6379-0013_1578-6379-0013.pdf}
% 	 %    \caption{}
% 		\label{fig:Ng2}
% 	 \end{subfigure}
	
% 	\caption{Initial development of cross-attention weights into the flipping.
% 	From top to bottom: after 2, 8, 10, and 12 epochs.}
% 	\label{fig:att-weights-evolution}
% \end{figure}

{
\setlength{\belowcaptionskip}{-8pt}
\begin{figure}
	\centering
	\includegraphics[width = 1.\columnwidth]{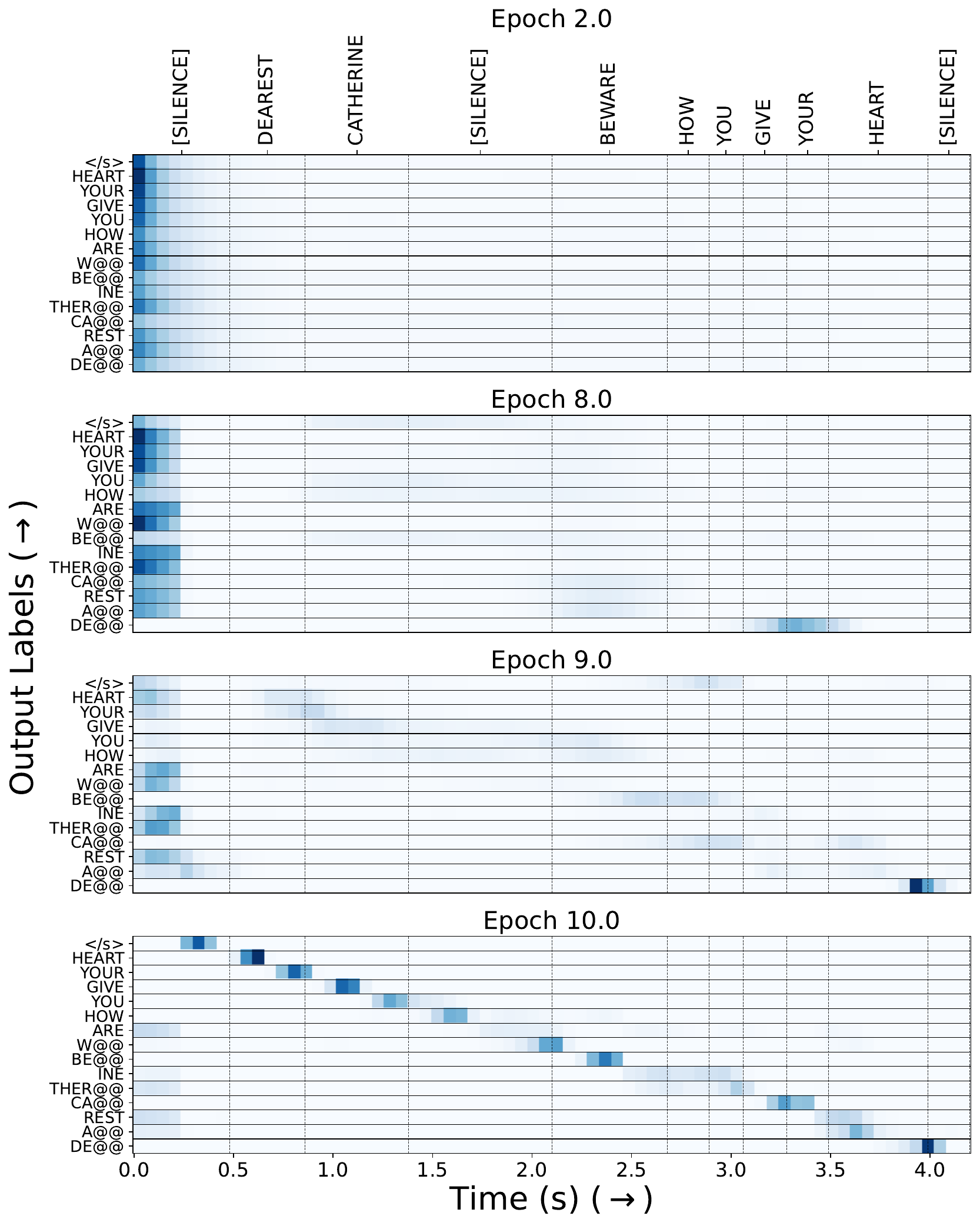}
	\caption{Development of cross-attention weights into the flipping behavior over the initial training epochs.}
	\label{fig:att-weights-evolution}
\end{figure}
}

\begin{figure}
	\centering
%	\begin{subfigure}[b]{0.4\textwidth}
	   \includegraphics[width=1\linewidth]{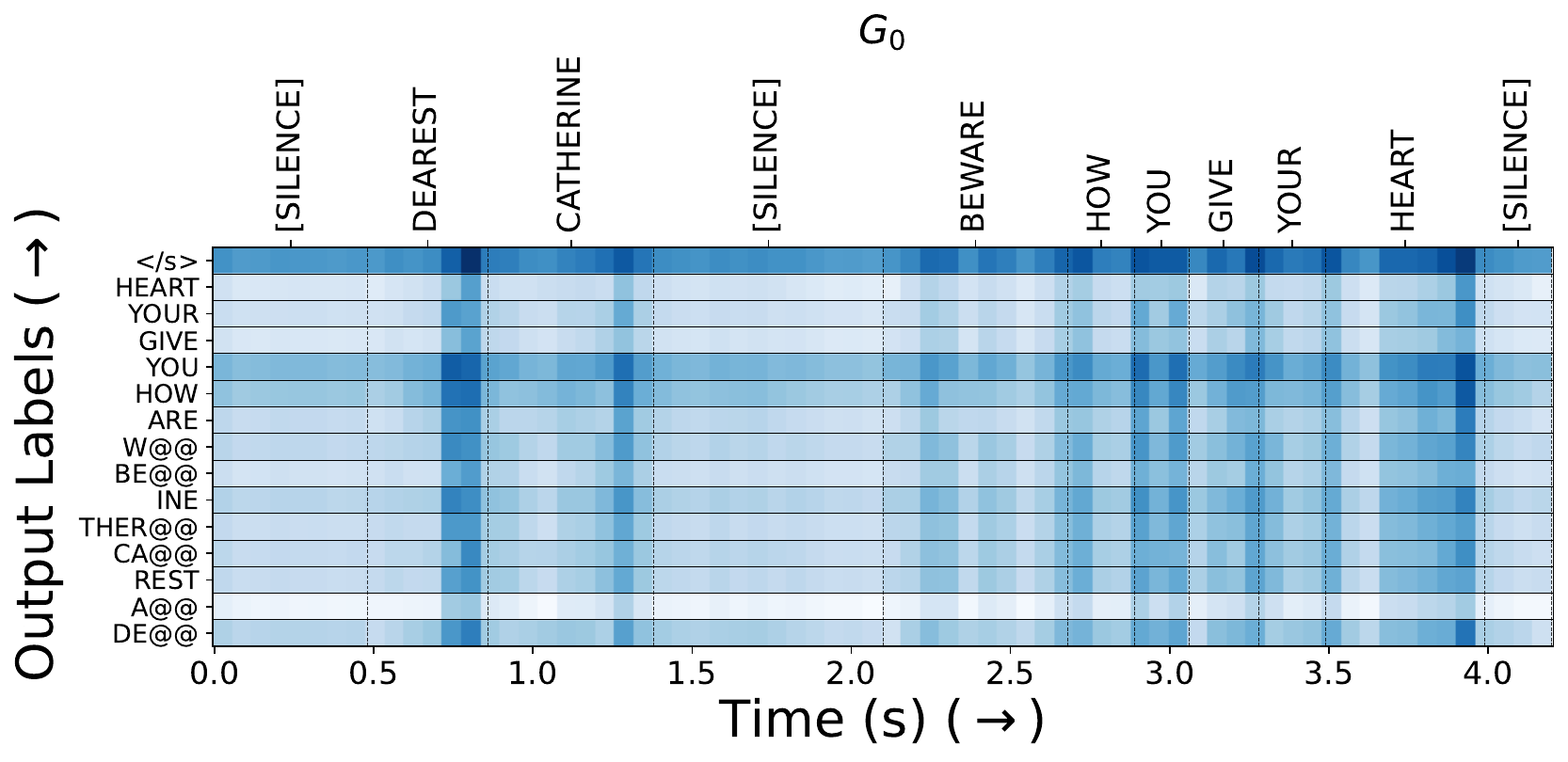}
%	   \caption{$G_0$: Conformer block 1 input.}
	   %\label{fig:Ng1} 
%	\end{subfigure}
	
%	\begin{subfigure}[b]{0.4\textwidth}
%	   \includegraphics[width=1\linewidth]{gradients/development/conformer-rand-4321/enc4_ep10_train-other-960_1578-6379-0013_1578-6379-0013.pdf}
%	   \caption{$G_4$: Conformer block 5 input.}
%	   \label{fig:Ng2}
%	\end{subfigure}
%
%	\begin{subfigure}[b]{0.4\textwidth}
%		\includegraphics[width=1\linewidth]{gradients/development/conformer-rand-4321/enc10_ep10_train-other-960_1578-6379-0013_1578-6379-0013.pdf}
%		\caption{$G_{10}$: Conformer block 11 input.}
%		\label{fig:Ng2}
%	 \end{subfigure}
	
	\caption{Showing $G_0$, i.e.~the logarithm of the $L_2$ norm of the gradients of the target label log probabilities
	w.r.t.~first Conformer block inputs very early in training (after 2 epochs).}
	\label{fig:gradient-wrt-enc}
\end{figure}
% i often observe that part of the att weights are flipped in the beginning but the model learns to correct this later on

In all of our experiments, the decoder cross-attention
initially only attends to the first few frames of the encoder output
(\Cref{fig:att-weights-evolution}).
We hypothesize that the choice of these frames is not due to their usefulness for predicting labels
but rather because they have distinct features, such as convolutional zero 
padding on the sequence boundaries and/or initial silence,
which makes it simple to attend to them.
The first frame is probably easier to recognize than the last,
as the last frame can have different padding due to the batching,
and also the sequences on LibriSpeech have slightly more silence at the beginning (260 ms vs. 230 ms on average).

We conducted an experiment where we force the model
to only attend to the center frame by setting the 
attention weight of this frame to one and all other weights to zero.
The initial CE losses of these models %(see \Cref{tab:initial-loss}),
are almost identical
(6.30 vs 6.29 after 1 epoch),
proving that the choice of frame for the initial cross-attention weights
is not important for label prediction early in training.
%This hypothesis is further supported by the fact that the output
%of the encoder does not contain any distinctive features in the beginning.
%as can be seen in \Cref{fig:enc1-sim} (\todo{} broken ref?).
%Instead, we think that the model chooses these first few frames
%because they have something in common for every sequence: silence. (\todo{} explain more)

To see how the model utilizes frames from individual encoder layers,
we examine the gradients of the target label $\overline{a}_s$ log probabilities
w.r.t.~the corresponding encoder layer input or output $h_i$
or the convolutional frontend input $h_{-1} = x$.
This gives us a $D$-dimensional vector per target label position $s$ and frame $t$,
which we reduce to a scalar by taking the logarithm of its $L_2$ norm:
\begin{equation}
{G_i}_{s,t} = \log \left\Vert \nabla_{{h_i}_t} \log p(\overline{a}_s \mid \overline{a}_1^{s-1}, h_1^T) \right\Vert_2 .
\label{eq:grad}
\end{equation}

Early in training (after 2 epochs), $G_0$ (\Cref{fig:gradient-wrt-enc}) shows
%where we can observe that
%in higher layers,
%the gradients are focused on the initial frames while,
%in middle layers, the gradients are still spread over the 
%whole sequence and,
%in lower layers,
that
the gradients are more focused
on individual label frames and less focused on the silence frames. 
For some labels, we can see that, in this early stage in training,
the gradients over all input frames are stronger,
meaning the encoder output is more important for those labels than for others.
It is specifically more pronounced for the labels "You" and \texttt{EOS}.

\subsection{How is Time Reversal possible?}

{
\setlength{\belowcaptionskip}{-8pt}
\begin{figure}
	\centering
	\includegraphics[width = 0.9\columnwidth]{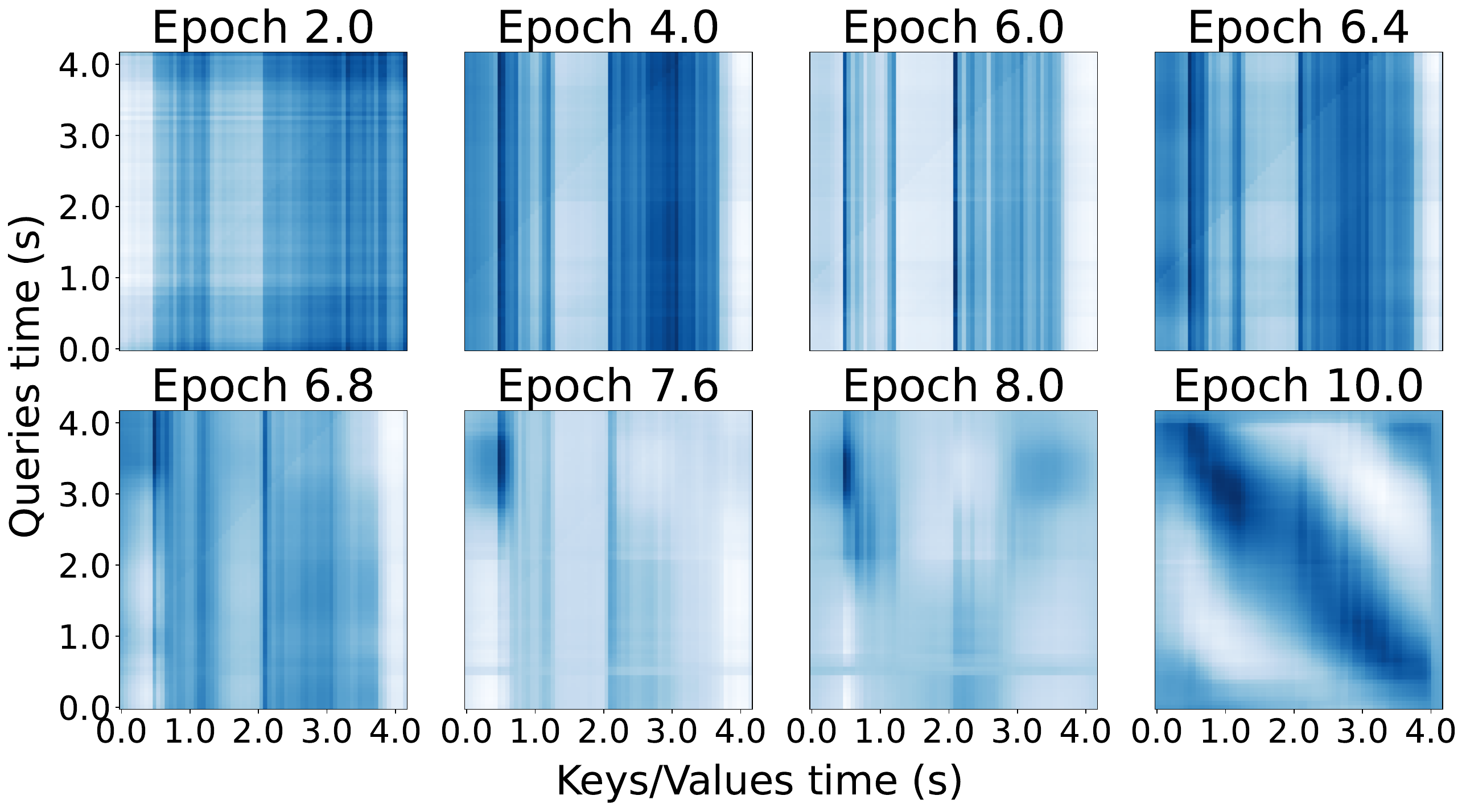}
	\caption{Self-attention energies averaged over the 8 heads of the 10th Conformer block for initial epochs.
	After this, all further layers are flipped.}
	\label{fig:enc9-flipped-self-att-ep12}
	\label{fig:flipped-self-att}
\end{figure}
}

% \begin{figure}
% 	\centering
% 	\includegraphics[width = 0.9\columnwidth]{self-att-energies/conformer-rand-4321/enc8_ep60_train-other-960_1578-6379-0013_1578-6379-0013.pdf}
% 	\caption{Self-attention energies of the 8 heads of the 8th Conformer block after 12 epochs.
% 	Here, self-attention already shifts information from first frames to last frames,
% 	but the residual connection still carries the non-flipped information}
% 	\label{fig:enc8-self-att-ep12}
% \end{figure}

% \begin{figure}
% 	\centering
% 	\includegraphics[width = 0.9\columnwidth]{self-att-energies/conformer-rand-4321/enc9_ep60_train-other-960_1578-6379-0013_1578-6379-0013.pdf}
% 	\caption{Self-attention energies of the 8 heads of the 9th Conformer block after 12 epochs.
% 	After this, all further layers are flipped.}
% 	\label{fig:enc9-flipped-self-att-ep12}
% 	\label{fig:flipped-self-att}
% \end{figure}

% \begin{figure}
% 	\centering
% 	\includegraphics[width = 0.9\columnwidth]{gradients/development/conformer-rand-4321/enc8_ep60_train-other-960_1578-6379-0013_1578-6379-0013.pdf}
% 	\caption{Gradients $G_8$ wrt.~block 8 output after 12 epochs. Here, we have the crossing of information from the residual and the self-att.}
% 	\label{fig:grad-G8out-cross}
% \end{figure}

% \begin{figure}
% 	\centering
% 	\includegraphics[width = 0.9\columnwidth]{gradients/development/conformer-rand-4321/enc9_ep60_train-other-960_1578-6379-0013_1578-6379-0013.pdf}
% 	\caption{Gradients $G_9$ wrt.~block 9 output after 12 epochs. Here, only the flipped information is left.}
% 	\label{fig:grad-G9out-flipped}
% \end{figure}

\begin{figure}
	\centering
	\includegraphics[width = 1.\columnwidth]{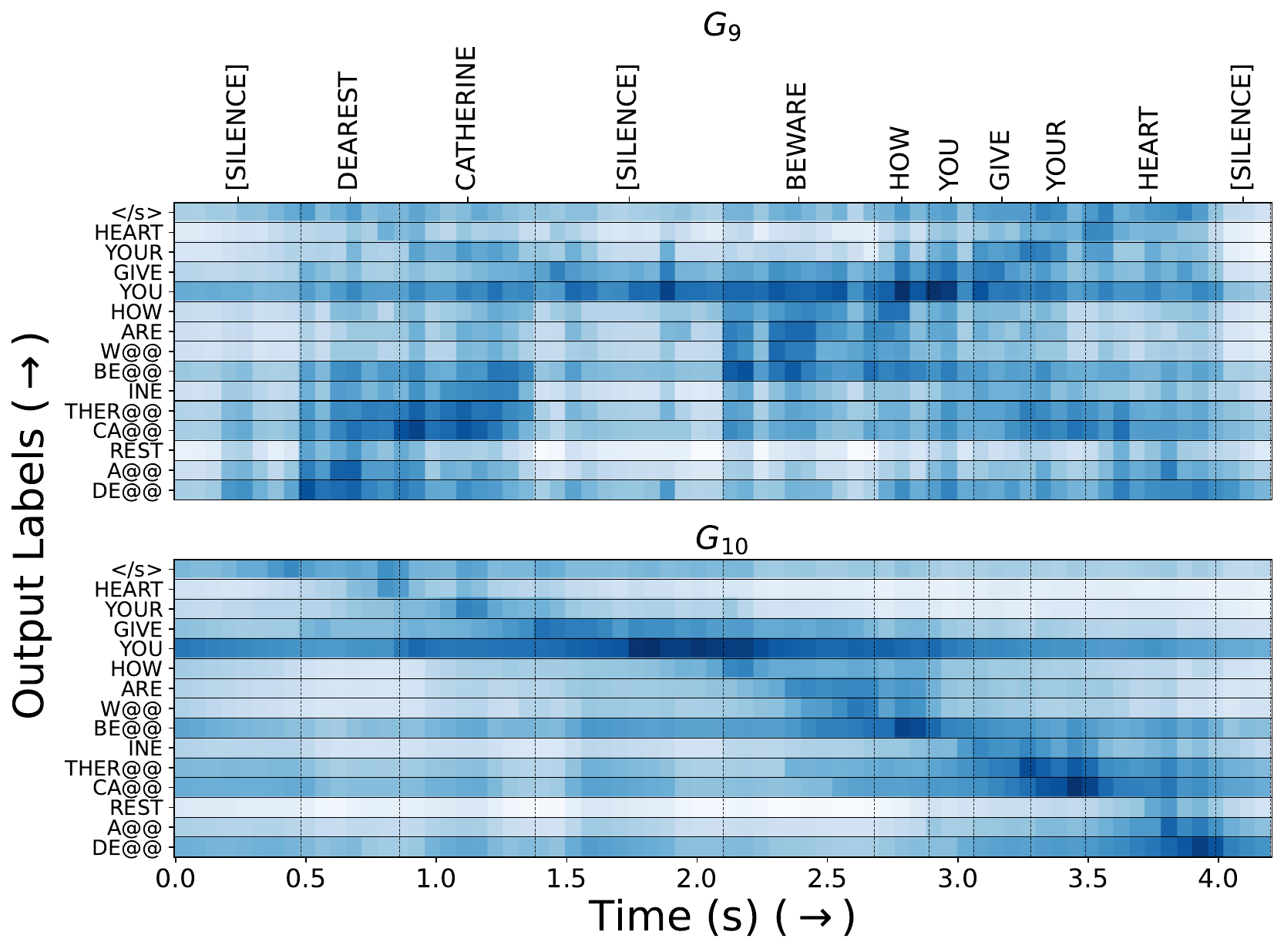}
	\caption{Gradients $G_9$ and $G_{10}$ w.r.t.~the output of blocks 9 and 10 after 12 epochs. For $G_9$, we have the crossing of information from the residual and the self-attention.
	In $G_{10}$, only the flipped information is left.}
	\label{fig:grad-G9out-cross-G10out-flipped}
\end{figure}

% -- I don't think the activation plots are really useful. The most important information is just the absolute magnitude (L2 norm) of the activations, but that's not visible in the figures. That's just a single number you can put into the text.
%\begin{figure}
%	\centering
%	\includegraphics[width = 0.9\columnwidth]{layer-activations/conformer-rand-4321/enc9_ff1_ep60_train-other-960_1578-6379-0013_1578-6379-0013_.pdf}
%	\caption{Activations of first FF module of 9th Conformer layer after 12 epochs.}
%	\label{fig:flipped-self-att}
%\end{figure}
%
%\begin{figure}
%	\centering
%	\includegraphics[width = 0.9\columnwidth]{layer-activations/conformer-rand-4321/enc9_mhsa_ep60_train-other-960_1578-6379-0013_1578-6379-0013_.pdf}
%	\caption{Activations of MHSA module of 9th Conformer layer after 12 epochs.}
%	\label{fig:flipped-self-att}
%\end{figure}

The flipping can only occur in the self-attention
(\Cref{fig:flipped-self-att}).
But what about the residual connection?
In every Conformer block, there is a final layernorm layer,
so there is no direct residual path from the input to the output of the encoder.
There is one Conformer block where the time reversal occurs.
In the flipping block, the self-attention module
has by far the largest activations in magnitude
% /u/schmitt/experiments/03-09-24_aed_flipped_encoder/work/i6_core/returnn/forward/ReturnnForwardJobV2.mmfDMb4GUqOm/work/enc-8/intermediate_layer_activations/activation_magnitude.txt
% in enc-8, the ff1 output is very low (3), while when it adds the residual (ff1_out), it gets higher (23),
% so the input to the block is already somewhat higher, and ff1 just doesn’t do much. then mhsa dominates with 60,
% and the others (conv and ff2) don’t change much anymore
(mean $L_2$ norm 60 for some example sequence),
so that the residual connection (norm 23)
from the input of this block and from the first feed-forward module
do not have much effect anymore.
The remaining convolution module and second feed-forward module
also do not add much (norm 13 and 15 respectively).
The final layernorm then removes all the original frame-wise information
and only the reverse order is kept.

Flipping occurs in Conformer block 10
and not in the other blocks (\Cref{fig:grad-G9out-cross-G10out-flipped}).
The gradients $G_9$ (\Cref{fig:grad-G9out-cross-G10out-flipped})
show that there is still some information remaining from the inner residual connection
in block 9.

We do not expect that such flipping is easy to perform by an encoder
where there is a residual connection from input to output
like in the standard Transformer \cite{vaswani2017transformer}.
%Without the final layernorm, the time reversal does not happen
% -- did not converge... not sure we need to mention this at all then... no real information except that we didn't correctly tune it

\subsection{Reasons for Time Reversal}

%\begin{figure}
%	\centering
%	\includegraphics[width = 1.0\columnwidth]{gradient-graph/conformer-rand-4321/enc8-9-10_ep60_train-other-960_1578-6379-0013_1578-6379-0013.pdf}
%	\caption{Gradient graph of encoder layers 8, 9 and 10 after 12 epochs. (\todo{} not sure if we should leave this figure in here, but maybe it helps for analyzing now)}
%	\label{fig:gradient-graph}
%\end{figure}

\emph{1)} The cross-attention initially only attends to the first few encoder frames
for all decoder frames (\Cref{fig:att-weights-evolution}, epoch 2)
as discussed before (\Cref{sec:initial-dev-cross-att}).

\emph{2)} The decoder acts initially like a language model and can learn independently from the encoder.
To slightly improve the prediction perplexity,
having some information of the encoder is useful,
and that is where it uses the fixed cross-attention to the first frame.
% -- Note, we had this earlier explanation first: I thought we saw this in some self-att. But maybe I misremembered. Now we do not see it like that. It does not attend to the last frame. Instead, it's more spread among multiple frames, more global.
%The inner self-attention in the encoder
%will prefer similar frames (e.g. left/right boundary, maybe silence)
%very early in training,
%i.e.~the first frames can easily attend to the end of the sequence.
%Now we ask the question: What information is more helpful for the decoder,
%the beginning or the end of the sequence?
%We measure the average self-information ($-\log_2(p(a_s))$
%over the training sequences
%per label position $s$
%based on label unigram probabilities ($p(a)$)
%and see that the first BPE label has lower self-information than the last BPE label on average (8.07 vs 9.71).
%Thus, it is more helpful for the decoder to get information from the last frame.
%The cross-attention already attends to the first frame,
%thus the encoder self-attention strengthens the attention from first to last frame.
%This can be seen in (\todo{} wrong...). %\Cref{fig:att-weights-evolution}.
The first encoder frame attends globally to more informative frames of lower layers
(\Cref{fig:flipped-self-att}, epoch 2 to 6.4)
as this is most useful at this stage to collect global information from all labels.
Attending globally to multiple frames is also more informative
than just first label or another single label
for predicting the whole sequence.

\emph{3)} To further improve the sequence prediction perplexity,
the cross-attention learns to focus on another frame,
which happens to be some random frame towards the end (\Cref{fig:att-weights-evolution}, epoch 8).
%It cannot use the first frame because that is already used to pass the global information.
The next most easily usable information is to know about the first label in the sequence%
\footnote{The last label is probably more difficult for the decoder to handle,
because it must also learn when that last label occurs.
It’s very easy to know that it must predict the first label.}.
Thus this encoder frame uses self-attention to attend to the frames of the first label
(\Cref{fig:flipped-self-att}, epoch 6.8 to 8),
where the first label is originally located in lower layers.

\emph{4)} The next labels follow, one after the other.
For the cross-attention, it is easier to choose some position right next to the previous position%
\footnote{Due to positional information,
when it has attended close to it in the previous decoder frame,
it should be easy to identify the next closest label in the encoder frames.}.
%Also, not even pos info, the encoder can prepare some special info like “here i’m the next place to attend to after you have attended to this other pos before.
So this will lead to the flipping
(\Cref{fig:att-weights-evolution} and \Cref{fig:flipped-self-att}, epoch 8 to 10).

%However, in order to correctly predict all labels,
%the model needs to utilize information from all frames.
%Therefore, by using the self-attention mechanism,
%the model can move information from other frames of the lower layers to these initial frames of the higher layers.
%This is reflected in the self-attention energies shown in \Cref{fig:enc8-self-att-ep12},
%where initial frames of heads 0, 1, 3, 6 and 7 attend more to the last frames.

% -- this here is again part of the older explanation, updated now.
%As the training progresses, the decoder can benefit from more information,
%and the cross-attention slowly learns to also look at other frames.
%As the first frame contains already the info on the last frame (flipped),
%the more right frames develop a connection to the first frame,
%as can be seen in \Cref{fig:flipped-self-att}.
%This evolves into the flipped behavior.
%(\todo{} update argumentation, it might happens simultaneously, due to symmetric similarity)

Vocabulary and average sequence length influence those training dynamics.
Specifically, skipping sequences longer than 75 labels during training
prevents flipping,
both for BPE1k and BPE10k.
For BPE1k, the output sequences are naturally longer.

The training dynamics are stochastic and depend on the random initialization.
Across 13 experiments with different random seeds,
the flipping happens in 12 cases (for BPE1k).
In 2 out of the 12 flipping cases, we also observed some shuffling of segments.

\subsection{Measures to Avoid Time Reversal}

\subsubsection{Use CTC auxiliary loss \cite{hori2017attctc}}
CTC enforces monotonic alignments between input frames and output labels,
so the encoder output can not be flipped.
We never observed flipping in experiments using CTC auxiliary loss.
This is commonly used
(by default in ESPnet \cite{watanabe2018espnet}
and many RETURNN setups \cite{zeyer2018interspeech,zeineldeen2024:ChunkedAED}),
which is why the flipping was maybe not observed before.

\subsubsection{Disabling self-attention in the beginning}

Flipping occurs in the self-attention.
We can fix the self-attention weights to be the identity matrix for the first epoch
and then only later use learned attention weights.
This means that we only use the linear transformation for the values
in the self-attention module.
This experiment shows faster convergence
(after 12 epochs: baseline reaches 0.83 CE loss,
baseline with CTC aux loss reaches 0.56 CE,
disabling self-attention reaches 0.45 CE),
and also no flipping occurs.
%To measure the speed of convergence,
%we compare how long it takes different models to reach
%a cross-entropy loss of below 1.0 on the dev set.
%Our baseline needs 11.6 epochs
%and disabling the self-attention for the first epoch reduces this to 7.4 epochs.
%In contrast,
%when leaving the self-attention on
%but using an additional CTC loss on top of some encoder layers
%(which is often done to speed up the convergence of global AED models),
%the model needs 6.8 epochs.

\subsubsection{Hard attention on center frame}

We argued before that the initial focused cross-attention to the first frame
might lead to the flipping.
We tried forcing the cross-attention weights to the center of the encoder output sequence
for the first full epoch.
In this experiment, no flipping did occur.\footnote{But we might need more experiments to really be sure.}

%\subsubsection{Cut off initial silence?}
%(\todo{} what is the result? flipping? focus still on beginning in early training?)

% skip this: we need to save space
%\subsubsection{Use Conformer input as attention keys?}
%
%The cross-attention uses the final encoder output both for the keys and the values.
%As the encoder input cannot be flipped yet,
%we thought we could use the encoder input for the keys of the cross-attention.
%That will force the model that all information in the final encoder output
%stay in the same frame as they are in the input,
%i.e.~it cannot flip the sequence.
%
%This did not work: after 130 sub-epoch (multi-gpu), we get 51.9\% WER.

\section{Alignments from Model Input Gradients}

\begin{figure}
	\centering
	\includegraphics[width = 1.\columnwidth]{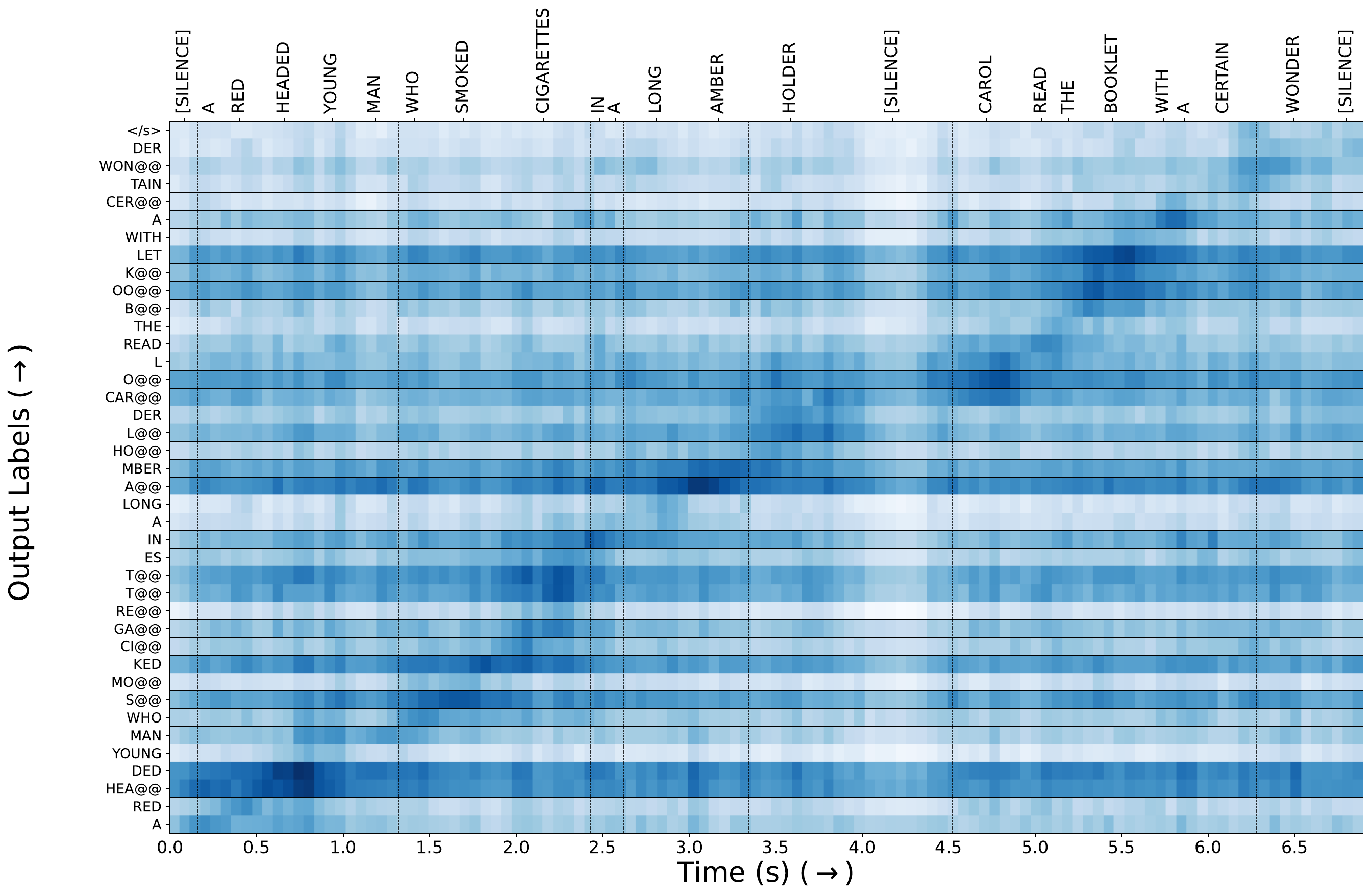}
	\caption{Gradients $G_0$ w.r.t.~Conformer input after 12 epochs.
	Alignment path is visible.
	%Used to find best fixed label-frame-position path.
	}
	\label{fig:grad-conformer-input}
\end{figure}

As shown in \Cref{fig:grad-conformer-input},
the \emph{gradients of the log label probabilities w.r.t.~the Conformer input} $G_0$
show an alignment between output labels and input frames.
We were asking the question:
Can we use those gradients
to estimate an alignment,
i.e.~the best alignment path (boundaries/positions of each label and word)?
This has been done before using the attention weights \cite{whisper_paper}
but that can be problematic due to multiple cross-attentions
(e.g.~Transformer with multiple layers and heads)
or because the encoder shifts or transforms the input,
as our work shows here.
When the model has reasonable performance,
the gradients w.r.t.~the model input cannot have those artifacts
like flipping or shifting.

\newcommand{\blank}{\ensuremath{\epsilon}}

We use either $G_0$ (\Cref{eq:grad}) to get an alignment with 60ms frame shift
or $G_{-1}$ (before the conv.~frontend) with 10ms frame shift.
We allow any number of \blank{} (blank) labels between any of the real labels,
we allow the real label to be repeated multiple times over time,
and we exclude the final \eos{} label.
This is very similar to the CTC label topology
except that we do not enforce an \blank{} between two equal labels.
We search for an allowed state sequences $r_1^T : a_1^S$
for state indices $r_t \in \{1,\dots,2 \cdot S - 1 \}$ corresponding to states $Y = (\blank, 1, \blank, 2, \dots, S-1, \blank)$
which maximizes
\begin{equation}
\operatorname{GradScore}(r_1^T) = \sum_{t=1}^T \operatorname{GradScore}(r_t)
\end{equation}
\begin{equation}
\operatorname{GradScore}(r_t) =
\begin{cases}
\log\operatorname{softmax}_{\overline{t}}(G_i)_{Y_{r_t},t}, & Y_{r_t} \ne \blank, \\
\gamma_{\blank}, & Y_{r_t} = \blank
\end{cases}
\end{equation}
for some fixed blank score $\gamma_{\blank}$ which is a hyperparameter%
\footnote{We use $\gamma_{\blank} = -4$ for the 60ms shift ($G_0$)
and $\gamma_{\blank} = -6$ for the 10ms shift ($G_{-1}$).}.
The best $r_1^T$ can be found via dynamic programming.
We obtain the final alignment label sequence $y_1^T$ with
$y_t = \begin{cases} a_{Y_{r_t}}, & Y_{r_t} \ne \blank, \\ \blank, & Y_{r_t} = \blank \end{cases}$.

We measure the time-stamp-error (TSE)
\cite{Zhang-2021-LatticefreeBoostedMMI,Raissi-2023-HMMVsCTC,Raissi2024:AlignQuality}
of word boundaries,
i.e.~the mean absolute distance (in milliseconds) of word start and end positions
against a reference GMM alignment, irrespective of the silence.
Additionally, we also compute the TSE w.r.t.~the word center positions,
which might be a better metric for peaky alignments like CTC \cite{zeyer2021peakyctc}.
We summarize our findings in \Cref{tab:align-quality}.
Our method still performs worse in terms of TSE
compared to a well tuned phoneme-level CTC model from earlier work \cite{Raissi2024:AlignQuality},
however we improve over all our BPE-based CTC alignments by far.
We also see that our method still works even when the encoder flips the sequence.

%\cite{Raissi2024:AlignQuality}: 38ms TSE on LS train (full train - we only use a random 10h subset), using phoneme-level CTC.
\begin{table}
	\centering
	\caption{Alignment quality in terms of time-stamp-error (TSE) for word left/right boundaries and center positions
	against reference GMM alignment on a random 10h subset of the training data of LibriSpeech.
	$^{*}$: On the whole training data, and the computation is different:
	more consistent alignment due to same feature extraction as the GMM,
	while our models here use a different feature extraction.
	% -- the following was some wrong understanding, it should be consistent to us:
	%Also, when a 40ms frame overlaps a 10ms frame from the GMM reference alignment,
	%it's not counted for the TSE, while in our case here, we do count it.
	}
	\label{tab:align-quality}
	\setlength{\tabcolsep}{3pt}
	\footnotesize
	\begin{tabular}{|l|l|c|c|c|c|c|c|}
	\hline
	\multirow{3}{*}{\shortstack{Best\\Path\\Scores}}  & \multirow{3}{*}{Model}  & \multirow{3}{*}{\shortstack{Train\\Phase}} & \multirow{3}{*}{Flip} &  \multirow{3}{*}{\shortstack{Frame\\Shift\\{[ms]}}}  & \multirow{3}{*}{\shortstack{Label\\Units}} & \multicolumn{2}{c|}{TSE [ms]}   \\ \cline{7-8}
			  && &&&              &          \multirow{2}{*}{\shortstack{Left/\\Right}} & \multirow{2}{*}{Center}  \\
			   && &&&              & & \\ \hline
	\hline
	Probs. & CTC   \cite{Raissi2024:AlignQuality} & full & no & 40 & Phonemes & 38\rlap{$^{*}$}       & - \\ \hline
	\hline
	\multirow{2}{*}{Probs.} & \multirow{2}{*}{CTC} & \multirow{2}{*}{full} & \multirow{2}{*}{no} & \multirow{2}{*}{60} & BPE1k & 83 & 61 \\ \cline{6-8}
	 &  &&&& BPE10k & \llap{3}12 & \llap{3}06 \\ \hline
	\hline
	% Grads. & AED & half & no  & 60 & BPE1k & 66 & 54 \\ \hline
	\multirow{4}{*}{Grads.} & \multirow{4}{*}{AED} &
	% {"grad_name": "base-convMask-early61-10ms", "sm": True, "blank_score": -6},  # 55.7/42.5 (!!)
	\multirow{4}{*}{early} & \multirow{2}{*}{no} & 10 & \multirow{4}{*}{BPE1k} & 56 & 43 \\  \cline{5-5}\cline{7-8}
	% {"grad_name": "base-convMask-early61-60ms", "sm": True, "blank_score": -4},  # 61.0/50.3 (!)
	&&&  & 60 & & 61 & 50 \\ \cline{4-5}\cline{7-8}
	% {"grad_name": "base-flip-early141-10ms", "sm": True, "blank_score": -6},  # 72.3/53.4
	&&& \multirow{2}{*}{yes} & 10 && 72 & 53 \\ \cline{5-5}\cline{7-8}
	% {"grad_name": "base-flip-early141-60ms" (early), "sm": True, "blank_score": -4},  # 75.0/60.8
	&&&  & 60 && 75 & 61 \\
	\hline
	\end{tabular}
\end{table}

% Early vs mid: early can be better:
%         {"grad_name": "base-flip-early141-10ms", "sm": True, "blank_score": -6},  # 72.3/53.4
%        {"grad_name": "base-flip-mid646-10ms", "sm": True, "blank_score": -6},  # 91.6/65.9
%           {"grad_name": "base-flip-far406-10ms", "sm": True, "blank_score": -6},  # 101.3/73.7  (other model though...)
% But is this always the case? Also the case for non-flipped.
%
% could also add this to table:
% {"grad_name": "base-convMask-mid225-10ms", "sm": True, "blank_score": -6},  # 65.2/50.5
% (but this is half/midway only... the LR is also biggest there due to OCLR. not sure if maybe difficult to argue...)
%
% Flipped vs non-flipped: non-flipped a bit better (already in table)

% no future work so far
\section{Conclusions}

In this work, we have shown that the Conformer encoder
of an attention-based encoder-decoder model
is able to reverse the time dimension of the input sequence.
% addition from rebuttal
%\textcolor{Green}{
The flipping is favored heavily due to the Conformer 
structure with its missing residual connections, not using CTC (or other 
monotonicity enforcing methods), and having longer sequence lengths on the 
decoder side.
%}
% \textcolor{Red}{
% We studied which functionality of the Conformer
% makes this behavior possible
% and how the initial cross-attention weights of the decoder push the model to do so.
% }
Furthermore, we proposed several methods to avoid this flipping
such as disabling the self-attention in the beginning,
which also improves convergence.

We also showed that the gradients of the label log probabilities
w.r.t.~the encoder input frames can be used to obtain alignments,
even early in training, even when the encoder reverses the sequence,
and its time-stamp-errors are even better than a normal CTC forced alignment.

\normalsize
\section{Acknowledgments} \footnotesize

This work was partially supported by the project RESCALE within the
program \textit{AI Lighthouse Projects for the Environment, Climate,
Nature and Resources} funded by the Federal Ministry for the
Environment, Nature Conservation, Nuclear Safety and Consumer
Protection (BMUV), funding ID: 67KI32006A.

\bibliographystyle{IEEEtran}
\bibliography{IEEEabrv,refs}

\end{document}